\documentclass[letterpaper]{article}
\usepackage{graphicx}
\usepackage{dcolumn}
\usepackage{bm}
\usepackage[width=17.70cm]{geometry}

\begin{document}

\title{Structure-Activity Relationship Investigation of Some New Tetracyclines by Electronic Index Methodology}

\author{F. Sato$^{1}$, S. F. Braga$^{1}$, H. F. dos Santos$^{2}$, and D. S. Galv\~{a}o$^{1}$}

\maketitle

$^{1}$Instituto de F\'{\i}sica Gleb Wataghin, Universidade Estadual de Campinas - Unicamp, Campinas - SP CP6165, CEP 13081-970, Brazil.

$^{2}$NEQC (N\'{u}cleo de Estudos de Qu\'{\i}mica Computacional), Departamento de Qu\'{\i}mica, Instituto de Ci\^{e}ncias Exatas, Universidade Federal de Juiz de Fora, Campus Universit\'{a}rio, Martelos, Juiz de Fora, MG, 36.036-900, Brazil.

\begin{abstract}
Tetracyclines are an old class of molecules that constitute a broad-spectrum antibiotics. Since the first member of tetracycline family were isolated, the clinical importance of these compounds as therapeutic and prophylactic agents against a wide range of infections has stimulated efforts to define their mode of action as inhibitors of bacterial reproduction. We used three SAR methodologies for the analysis of biological activity of a set of 104 tetracycline compounds. Our calculation were carried out using the semi-empirical Austin Method One (AM1) and Parametric Method 3 (PM3). Electronic Indices Methodology (EIM), Principal Component Analysis (PCA) and Artificial Neural Networks (ANN) were applied to the classification of 14 old and 90 new proposed derivatives of tetracyclines. Our results make evident the importance of EIM descriptors in pattern recognition and also show that the EIM can be effectively used to predict the biological activity of Tetracyclines.
\end{abstract}



\section{Introduction}\label{intro}

Tetracyclines (TC) are a set of compounds used in human therapy against a broad-spectrum of microbial agents since 1947. They can be classified as low cost drugs with small side-effects being largely used by food industry and for therapy in animals and plants \cite{Levy1992}. TC present activity against Gram-positive and Gram-negative anaerobic and aerobic bacteria, mycrobacterium and several protozoan parasites \cite{Roberts1996}. Non-antibacterial effects as anti-inflammatory, immunosuppressive \cite{Humbert1991}, antioxidant \cite{Lertvorachon2005} and anticancer  \cite{Fife1998,Hidalgo2001} are also attributed to them.

The excessive use of TC during years helped resistant bacteria proliferation, limiting then their clinical use. More than 30 TC resistance determinants have been described  \cite{Levy1999,Roberts2005} and several different TC resistance determinants have been identified in E. coli \cite{Sengel2003,Levy1985,Jones1992}. This emergence of resistant pathogens is becoming an increasingly important problem and it motivates the search for new TC variations and derivatives  \cite{Chartone2005}. TC can be divided into two classes, the bacteriostatic typical TC and atypical TC, which are bactericidal and they exert their effects by promoting bacterial autolysis. Most probably due to their different modes of action, atypical TC have been shown to exert activity against some TC resistant bacteria \cite{Oliva1992,Halling2002}. 

The chemistry of TC in solution is quite complicated \cite{Cosentino2005} due to their ability to adopt different conformations, protonation states, and tautomeric forms, depending on the conditions investigated \cite{Othersen2003}. TC and its analogues undergo complex formation with a variety of the metal cations present in biological fluids \cite{Lambs1988,Lambs1989,Brion1986,Lambs1985} and this affects their chemical as well as conformational equilibrium in solutions. Theoretical and experimental investigations have contributed to the understanding of the interaction of TC with biological medium and their mode of action  \cite{Song2005,Baptiste2004,Nicolas2003,Santos1998,Santos2003,Santos2006}. Theoretical calculations have also been applied to investigations of new applications for old, new and hypothetical TC derivatives \cite{Baptiste2004}.

\begin{figure}
\caption{\label{fig1} Basic structure of Tetracycline.}
\includegraphics[width=8.0cm]{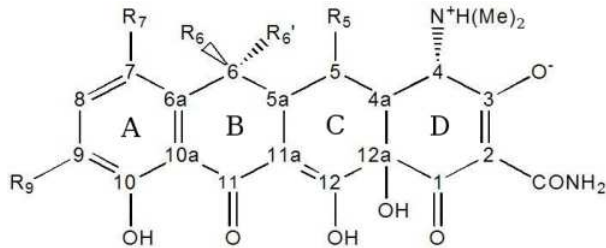}
\end{figure}

In this work we present an investigation of electronic and quantum features for a series of TC (Figure \ref{fig1}, Table \ref{tab1}) for which the biological activity against \textit{E. Coli} is known and also to new compounds \cite{peradejordi1971}. Our results showed that it is possible to directly correlate some quantum electronic descriptors to TC biological activity. This information is then used in the design of new compounds.

\begin{table*}
\caption{\label{tab1}Set of investigated Tetracyclines labelled accordingly to Figure \ref{fig1} and respective biological activity index \cite{peradejordi1971}. For compound 12 the substituintes on carbon 11 and 12 are interchanged and for compound 13 carbons 3 and 4 are linked to hydrogen atoms.}
\begin{tabular}{cccccccccc}

Molecule & \multicolumn{2}{c}{Compound} & \multicolumn{2}{c}{R$_{5}$} & R$_{6}$ & R$_{6\mbox{'}}$ & R$_{7}$ & R$_{9}$ & $K_{sens} \times10^{3}$ \\ \hline

01 & \multicolumn{2}{c}{Tetracycline} & \multicolumn{2}{c}{H} & OH & Me & H & H & 600.0 \\

02 & \multicolumn{2}{c}{7-NH$_{2}$-6-Demethyl-6-deoxytetracycline} & \multicolumn{2}{c}{H} & H & H & NH$_{2}$ & H & 190.0 \\

03 & \multicolumn{2}{c}{5-Hydroxytetracycline} & \multicolumn{2}{c}{OH} & OH & Me & H & H & 650.0 \\

04 & \multicolumn{2}{c}{9-NO$_{2}$-6-Demethyl-6-deoxytetracycline} & \multicolumn{2}{c}{H} & H & H & H & NO$_{2}$ & 145.0 \\

05 & \multicolumn{2}{c}{9-NH$_{2}$-6-Demethyl-6-deoxytetracycline} & \multicolumn{2}{c}{H} & H & H & H & NH$_{2}$ & 350.0 \\

06 & \multicolumn{2}{c}{7-Cl-6-Demethyltetracycline} & \multicolumn{2}{c}{H} & OH& H & Cl & H & 1100.0 \\

07 & \multicolumn{2}{c}{7-Cl-Tetracycline} & \multicolumn{2}{c}{H} & OH & Me & Cl & H & 950.0 \\

08 & \multicolumn{2}{c}{7-NO$_{2}$-6-Demethyl-6-deoxytetracycline} & \multicolumn{2}{c}{H} & H& H& NO$_{2}$& H& 1450.0 \\

09 & \multicolumn{2}{c}{9-N(CH$_{3})_{2}$-6-Demethyldeoxytetracycline} & \multicolumn{2}{c}{H} & H& H& H& N(Me)$_{2}$& 76.0 \\

10 & \multicolumn{2}{c}{6-Demethyl-6-deoxytetracycline} & \multicolumn{2}{c}{H} & H& H& H& H& 250.0 \\

11 & \multicolumn{2}{c}{6-Demethylamino-6-demethyldeoxytetracycline} & \multicolumn{2}{c}{H} & H& H& N(Me)$_{2}$& H& 250.0 \\

12& \multicolumn{2}{c}{5a,6-Anhydrotetracycline} & \multicolumn{2}{c}{H} & Me& -& H& H& 130.0 \\

13& \multicolumn{2}{c}{12a-Deoxytetracycline} & \multicolumn{2}{c}{H} & OH& Me& H& H& 8.5 \\

14 & \multicolumn{2}{c}{7-Br-6-Demethyl-6-deoxytetracycline} & \multicolumn{2}{c}{H} & H & H & Br & H & 70.0 \\

\end{tabular}
\end{table*}

\section{Methodology}\label{metodologia}

We started our investigation with a detailed search for the global minimum of energy for 14 TC derivatives with known biological activity (Figure \ref{fig1}, Table \ref{tab1}). In the geometrical search of each molecule all degrees of freedom have been allowed to change during optimization and special atention has been devoted to the $NH(Me)_{2}$ group bounded to carbon 4, which is known to influence TC solution and gas phase stability \cite{Othersen2003}. All the calculations have been performed with semiempirical Austim Method 1 (AM1) \cite{dewar1985} and Parametric Method 3 (PM3) \cite{steward1991a, steward1991b}. The AM1 and PM3 methods are widely used in the investigation of organic molecules with very good results \cite{zerner1991}. In this study one of the points we would like to evaluate is wheter the geometry and the correlation of theoretical properties with biological activity are dependent on the used semiempirical method. Molecules are studied in gas phase adopting the zwiterionic form, that is one of three TC basic forms and considered to be the active species in the biological medium. In zwiterionic form the total charge in molecule is equal zero but the groups bounded to the carbons 3 and 4 are charged and set to $O^{-}$ and $NH^{+}(Me)_{2}$, respectively \cite{peradejordi1971}. Although we can expect that Hartree-Fock methods such as AM1 and PM3 could have some difficulties in precisely discribing these forms, a systematic error is expected. Zwiterionic together with NH$^-$ ionized species are commonly present in aqueous solutions and it is believed to complex with ions metals as Ca$^{2+}$, Mg$^{2+}$, Fe$^{2+}$, Zn$^{2+}$, Al$^{3+}$ in biological medium, affecting the bioavailability and therapeutic  TC properties \cite{Siqueira1994,Machado1995,Mello1995}.

For the geometry optimization and properties calculation we have used the Chem2Pac package  \cite{cyrillo1999}. Once optimized, the conformations of minimun energy were the basis for theoretical properties and electronic descriptors calculations. We based our structure-activity relationship (SAR) investigation on the Electronic Indices Methodology (EIM) and its descriptors. EIM was initially developed to identify the carcinogenic activity of polycyclic aromatic hydrocarbons \cite{barone1996,santo1999} using very simple rules. The methodology uses only two descriptors called $\Delta$ and $\eta$. The $\Delta$ is defined as the energy separation of molecular levels as indicated in Eq.(\ref{deltah}).

\begin{footnotesize}
\begin{equation}\label{deltah}
\Delta = E_{level 1} - E_{level 2}
\end{equation}
\end{footnotesize}

The descriptor $\eta$, is based on the Local Density of States (LDOS), which is obtained as the Density of States (DOS) for a selected group of atomic orbitals or atoms. The $\eta$ is calculated as the LDOS difference between the most relevant molecular electronic levels (defined in the $\Delta$ calculation) over molecular regions identified as related to biological activity:

\begin{footnotesize}
\begin{equation}\label{etah}
\eta = 2\sum\limits_{a = a_1 }^{a = a_n } {(\vert C_{a (level 1)}\vert ^2 - \vert C_{a (level 2)}\vert ^2)}
\end{equation}
\end{footnotesize}

$C_a$ is the coeficient of atomic orbital from LCAO (Linear Combination of Atomic Orbitals) approximation. $a_1$ and $a_n$ refer to atoms composing the selected molecular region. $\Delta$ contains global molecular information (involving the eigenvalues) while $\eta$ contains local molecular information (LDOS) assumed to reflect the molecular environmental biochemical mechanisms (for details of EIM see refs.  \cite{barone1996,santo1999,braga2000,sbraga2004,cyrillo1999,katro2004,sbraga2003}).

With only these two descriptors EIM has classified with sucess, as active and inactive, molecules of differente classes of activity including steroid hormones \cite{braga2000,sbraga2004}, integrase inhibitors \cite{cyrillo1999}, antibiotics \cite{cyrillo1999}, and tumoral \cite{katro2004} and antitumoral compounds \cite{sbraga2003}. Usually the EIM rules involve critical values assuming the boolean form: IF $\Delta > (<) \Delta_{c}$ AND(OR) $\eta > (<) \eta_{c}$, the molecule will be active, otherwise it will be inactive. $\Delta_{c}$ and $\eta_{c}$ are critical values determined from EIM analysis. The critical values and EIM rules are automatically obtained from an exploratory scanning. In contrast to other SAR methods, EIM does not use training sets. 

In order to obtain the EIM parameters and classificatory rules, the common molecular skeleton of the group of compounds investigated is divided into regions. Available informations about important molecular regions to determine biological activity is taken into account in the choice of the regions to the LDOS calculations. The present study we divided TC into 6 regions:

\begin{enumerate}
\item{ring A and radicals;}
\item{ring B and radicals;}
\item{ring C and radicals;}
\item{ring D and radicals, exception to groups $CONH_{2}$ and $N^{+}H(Me)_{2}$;}
\item{radical $N^{+}H(Me)_{2}$;}
\item{radical $CONH_{2}$.} 
\end{enumerate}

The group of 14 molecules has been divided into two subgroups accordingly to their biological  activity. Our criterion to classify molecules as active or inactive was based on the experimental values of $K_{sens}$ index presented in Table \ref{tab1}. Molecules with $K_{sens}$ lower than 100 are classified as inactive and those with $K_{sens}$ greater than $100$ as actives. The $K_{sens}$ index represents the inhibition velocity constant based on cell growing velocity law  \cite{peradejordi1971}.

The importance of electronic descriptors was evaluated by comparing EIM results with those obtained with standard SAR methods, Principal Component Analysis (PCA) and Artificial Neural Networks (ANN).

The PCA method is an extremelly useful explorative tool to separate samples into groups accordingly to their similarity and also to select important descriptors to these separations. In a geometrical interpretation each descriptor is considered a vector in a N-dimensional space and PCA method constructs a new space where new vectors, called principal components (PC), are linear combinations of the original descriptors. In the PCs space the samples are mapped through scores and the original descriptors through loading. From the loading plots, the important descriptors can be identified and the correlation pattern among them can be established. These calculations were carried out with the \textit{Einsight} program \cite{einsight} 

The ANN analysis was carried out using the program package Perceptron-type Neural Network Simulator with back-propagation algorithm (PSDD) \cite{psdd1, psdd2}. PSDD consists of a three-layer perceptron network originally developed for drug design. This is a non-linear method and offers great advantage over conventional SAR methods for the cases where the target properties are not linearly dependent on the chosen parameters. In the ANN analysis we considered only the EIM descriptors to classify old and new proposed tetracyclines.

\section{Results and Discussions}\label{res}

\subsection{Structures}\label{str}

For the geometry optimization a conformational search was carried out considering the systematic variation of the dihedral angles related to the radicals bounded to the molecular skeleton. We initialy used the rotational step of 10 degrees to the OH, Me, NO$_{2}$, NH$_{2}$, CONH$_{2}$, and NH(Me)$_{2}$ groups in a rigid-rotor approach (one self consistent field calculation) using AM1 and PM3 hamiltonians. The ten best conformations (in terms of energy) were then selected and their geometries were fully optimized. All calculations (AM1 and PM3) have been performed with Mopac 6 implemented in Chem2Pac program \cite{cyrillo1999} using high quality numerical precision optimization criteria.

\begin{figure}
\caption{\label{barreira_all}Relative energy for the rotation of dihedral $Dh_(NH^{+}(Me)_2)$ calculated with AM1 method to molecule 01. The -180 and 180 degrees correpond to the same conformation, the called extended conformation. The twisted conformation are located around 35 degrees. Structures in the inset show details of these conformations.}
\includegraphics[width=8.0cm]{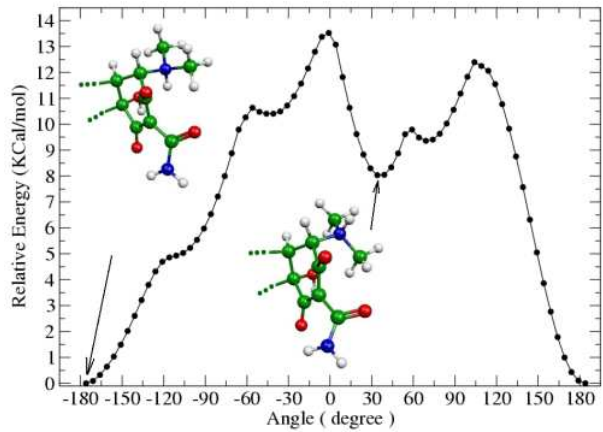}
\end{figure}

An important characteristic of TC geometries is the presence of $NH^{+}(Me)_2$ group bounded to carbon 4. For the most stable conformations we investigated in details the energy variation related to the rotation of this group. For this specific part of study we optimized the geometry of each conformation generated by a rotation of 5 degrees at the dihedral formed by the hydrogen and nitrogen atoms of $NH^{+}(Me)_2$ group, carbon 4, and hydrogen bound on carbon 4 ($Dh_{NH^{+}(Me)_2}$). The complete rotation of $360^{o}$ showed the existence of the $Dh_{NH^{+}(Me)_2}$ two local minima of energy. The structure at Dh=-180 degrees is named extended and the other at Dh=30 degrees twisted \cite{Santos1998,Santos2003,Othersen2003}. In Figure \ref{barreira_all} we present the result obtained to molecule 01, for all the other molecules the curve of energy is similar. The relative energy is calculated considering the lowest energy as zero of reference. As can be observed from Figure \ref{barreira_all} the lowest energy conformer corresponds to the dihedral $Dh_{NH^{+}(Me)_2}$ around 180 degrees. From the inset in the Figure \ref{barreira_all} we can observe that in this conformation the hidrogen of $NH^{+}(Me)_2$ is directed toward the OH group at C12a. The second local minimum conformation is observed for $Dh_{NH^{+}(Me)_2}$ around 35 degrees and in this case the hydrogen atom points to O3 moiety.

\begin{figure}
\caption{\label{ttc01}Molecule 01 optimized structure from AM1 calculation.}
\includegraphics[width=8.0cm]{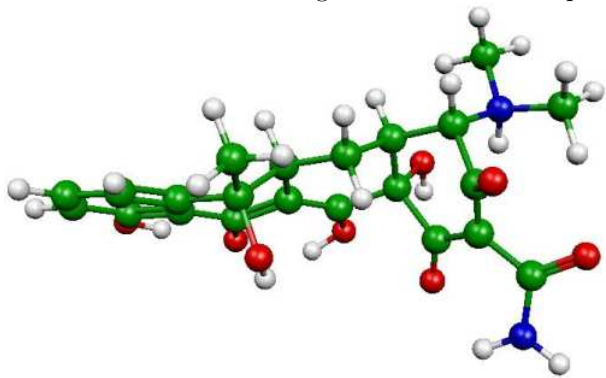}
\end{figure}

In order to better illustrate these conformations we show in Figure \ref{ttc01} the lowest energy conformer for molecule 01 in the extended conformation.

In Table \ref{conf} we present the main dihedral angles characterizing the TC rings from the global minimum of each compound calculated with AM1. Similar results were obtained with PM3 method.

As can be observed from the values of dihedrals 5-4a-12a-1 and 3-4-4a-12a (Table \ref{conf}) all the molecules have the A ring non co-planar with the plane formed by rings D, C, and B. The rings C and B present a small torsion induced by the hydrogen atom bounded to carbon 5a. Torsions angle values between D-C and C-B rings are shown in second and third columns, respectively.

\begin{table}
\caption{\label{conf} Main latex article templatedihedral angles (in degrees) for optimized TC structures with AM1 method. The second, third, and fourth columns show the planarities between rings D-C, C-B, and B-A, respectively. The fifth column indicates A ring angle torsions values. Labelling accordingly to Figure \ref{fig1}.}
\begin{tabular}{ccccc}
Molecule& 7-6a-10a-11& 6-5a-11a-12& 5-4a-12a-1& 3-4-4a-12a  \\ \hline
01&  -178.7& -138.5& -78.9& 60.6 \\
02&  -175.7& -147.3& -84.8& 60.1 \\
03&  -171.4& -121.6& -66.4& 22.1 \\
04&  -178.1& -139.1& -78.6& 62.4 \\
05&  -178.0& -139.7& -79.5& 62.4 \\
06&  -179.7& -140.9& -79.9& 60.5 \\
07&  -165.7& -141.7& -82.2& 62.7 \\
08&  -174.5& -140.9& -80.1& 60.7 \\
09&  -174.6& -140.0& -79.8& 61.0 \\
10&  -178.9& -139.4& -79.2& 62.4 \\
11&  -176.7& -139.8& -79.9& 61.0 \\
12&   174.1&  178.8& -72.3& 59.8 \\
13&  -177.9& -136.7& -72.2& 54.3 \\
14&  -177.8& -140.3& -79.9& 60.9 \\
\end{tabular}
\end{table}

\subsection{The EIM Method and PCA Analysis}\label{eim}

From the EIM analysis we obtained the $\Delta$ for pairs of frontier orbitals (Highest Occupied Molecular Orbital-HOMO, Lowest Unoccupied Molecular Orbital- LUMO and next closest levels) and the $\eta$ data for all the regions listed in the methodology section. We observed that the best results for both AM1 and PM3 methods are obtained for $\Delta H$ (Eq. \ref{delta2}) and $\eta H$ (Eq. \ref{eta2}) calculated over the region 6 (radical CONH$_{2}$ bounded to carbon 2). In this case the equations \ref{deltah} and \ref{etah} became:

\begin{footnotesize}
\begin{equation}\label{delta2}
\Delta = E_{HOMO} - E_{HOMO-1}
\end{equation}
\end{footnotesize}

\begin{footnotesize}
\begin{equation}\label{eta2}
\eta = 2\sum\limits_{a = a_1 }^{a = a_n } {(\vert C_{a
(HOMO)}\vert ^2 - \vert C_{a (HOMO-1)}\vert ^2)}
\end{equation}
\end{footnotesize}

\begin{table}
\caption{\label{tab2}EIM descriptors calculated from both AM1 and PM3 methods. Molecules are presented according to their $K_{sens}$ values \cite{peradejordi1971}.}
\begin{tabular}{ccccccc}
& & \multicolumn{2}{c}{ AM1 } & & \multicolumn{2}{c}{ PM3 }  \\ \hline

Molecule& $K_{sens}\times 10^{3}$& $\Delta $H& $\eta $H& & $\Delta $H& $\eta $H \\

13&    8.5&  0.98&  1.42&  & 0.64& -0.56 \\
14&   70.0&  0.89&  1.39 & \raisebox{-3.00ex}[0cm][0cm]{Inactive} & 0.93 & -0.56 \\
09&   76.0&  0.76&  1.39&  & 0.59& -0.63 \\

\multicolumn{7}{c}{}  \\

12&  130.0&  0.18&  1.26&  & 0.282&  1.23 \\
04&  145.0&  0.94& -0.59&  & 0.465& -0.72 \\
02&  190.0&  0.11& -1.38&  & 0.194& -1.28 \\
10&  250.0&  0.86&  1.33& \raisebox{-15.00ex}[0cm][0cm]{Active}& 0.50& -0.71 \\
11&  250.0&  0.39&  1.39&  & 0.599& -0.63 \\
05&  350.0&  0.12&  1.33&  & 0.311&  1.29 \\
01&  600.0&  0.99&  1.38&  & 0.581& -0.63 \\
03&  650.0&  1.00&  1.38&  & 0.893& -0.53 \\
07&  950.0&  0.81&  1.33&  & 0.575& -0.63 \\
06& 1100.0&  0.94&  1.34&  & 0.465& -0.71 \\
08& 1450.0&  1.18& -0.26&  & 0.554& -0.63 \\
\end{tabular}
\end{table}

The EIM parameter values are presented in Table \ref{tab2}. In order to build the EIM rules we determine the critical values of the descriptors $\Delta H_c$ and $\eta H_c$. The classificatory rules are constructed trying to reproduce the experimental classes indicated in section \ref{metodologia} with relation to the $K_{sens}$ values. For AM1 (Eq.\ref{eim-am1}) and PM3 (Eq. \ref{eim-pm3}) methods the following the rules have been obtained:

\begin{footnotesize}

\begin{equation}\label{eim-am1}
\mbox{IF } \Delta H \ge 0.70 \mbox{ AND } \eta H \ge 1.39 \mbox{ THEN molecule is inactive}
\end{equation}

\begin{equation}\label{eim-pm3}
\mbox{IF } \Delta H \le 0.58 \mbox{ OR } \eta H \ge -0.54 \mbox{ THEN molecule is active}
\end{equation}

\end{footnotesize}

These crittical values (in equations \ref{eim-am1} and \ref{eim-pm3}) were obtained from an automatic screening search that determine the best statistical relevance of these numbers.

With these two descriptors and EIM simple rules for AM1 data it was possible to reproduce the experimental activity (active or inactive) with 100\% of accurary. For PM3 calculation the rules achieved $\sim$93\% of accuracy, only molecule 11 was incorrectly classified.

The EIM results were then contrasted against the PCA methodology. For PCA analysis we considered a set of 25 parameters where electronic, physicochemical and stereochemical ones were included:

\begin{itemize}
\item{$\Delta H$}
\item{$\eta H$}
\item{$\Delta L$}
\item{$\eta L$}
\item{Energy$_{HOMO}$}
\item{Energy$_{HOMO-1}$}
\item{Energy$_{LUMO}$}
\item{Energy$_{LUMO+1}$}
\item{LDOS (values from eigenstates HOMO, HOMO-1, LUMO, and LUMO+1)}
\item{heat of formation}
\item{surface area}
\item{molecular volume}
\item{hydratation energy}
\item{Log P}
\item{refractivity}
\item{polarizability}
\item{molecular mass}
\item{electronic energy}
\item{nuclear repulsion}
\item{dipolar momentum}
\item{ionization potential}
\item{hardness.}
\end{itemize}

Physicochemical variables frequently used in QSAR analysis have also been included in the calculations to evaluate the relevance of EIM descriptors. All the parameters were obtained from AM1 and PM3 calculations and PCA calculations have been carried out separatedly for them. 

\begin{figure}
   \caption{\label{fig2}PCA - Scores from AM1 data.}
   \includegraphics[width=8.0cm]{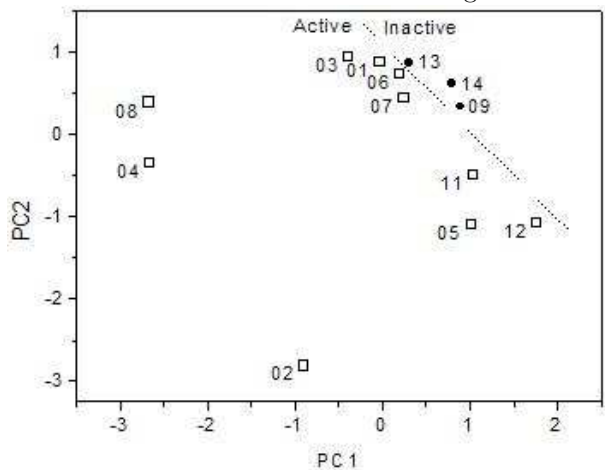}
\end{figure}

\begin{figure}
   \caption{\label{fig3}PCA - Score from PM3 data.}
   \includegraphics[width=8.0cm]{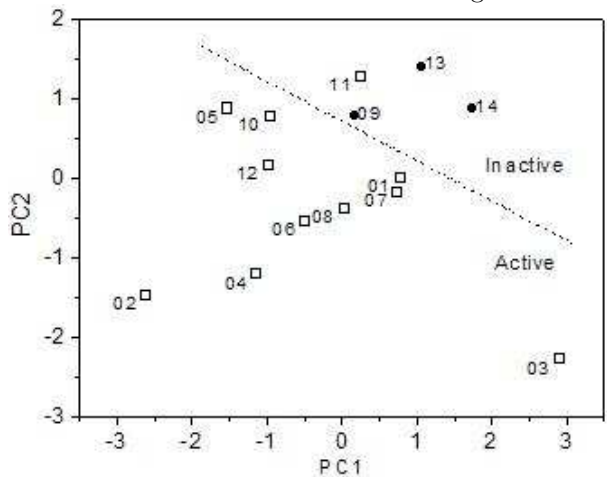}
\end{figure}

The data have been auto-scaled prior PCA analysis and we focused our attention on the two first principal components. The best results separating active and inactive compounds are presented in Figs. \ref{fig2} and \ref{fig3} to AM1 and PM3 descriptors, respectively.

Molecules in PCs planes are separated accordingly to the equations: 

\begin{footnotesize}
AM1
\begin{equation}\label{pc1-am1}
PC1=-0.36(\Delta H) +0.58(\eta H) +0.74(LogP)
\end{equation}

\begin{equation}\label{pc2-am1}
PC2= +0.81(\Delta H) +0.58(\eta H) -0.06(LogP)
\end{equation}

PM3
\begin{equation}\label{pc1-pm3}
PC1 = +0.08(HOMO) +0.66(\Delta H) -0.42(HF) +0.62(\eta H)
\end{equation}

\begin{equation}\label{pc2-pm3}
PC2 = +0.82(HOMO) +0.04(\Delta H) +0.54(HF) +0.21(\eta H)
\end{equation}
\end{footnotesize}

For AM1 and PM3 the PC1 vs PC2 plane accumulates a total variance of 87.26\% and  80.71\%, respectively.

From Figures \ref{fig2} and \ref{fig3} and equations \ref{pc1-am1}, \ref{pc2-am1}, \ref{pc1-pm3}, and \ref{pc2-pm3} we can see which variables are statistically more relevant among all the others in the set investigated. With exception of LogP all other variables are electronic and related to the EIM descriptors. Its is worth mentioning that during variables selection all variables were equally considered, no parameter scalling was used. In Fig. \ref{fig2} a line delimiting the two classes of compounds was traced. For this case the molecules  tend to agglomerate and the separation is not so clear. The results for PM3 data (Figure \ref{fig3}) show molecules more disperse and active and inactive groups separated. As already observed from EIM results, for the PM3 hamiltonian the molecule 11 was incorrectly classified as inactive. 

These EIM and PCA results can be used together as an aplicable and fast tool in the classification of new Tetracyclines. In this sense we propose a set of 90 new hypothetical TC to what experimental data is unknown. We then used the EIM rules from the available experimental data to classify (active/inactive) these new hypotetical structures.

\subsection{New Proposed Tetracyclines}\label{new}

\begin{figure}
\caption{\label{fig4}Structures of new proposed tetracyclines - $T_{\alpha}$}
\includegraphics[width=8.0cm]{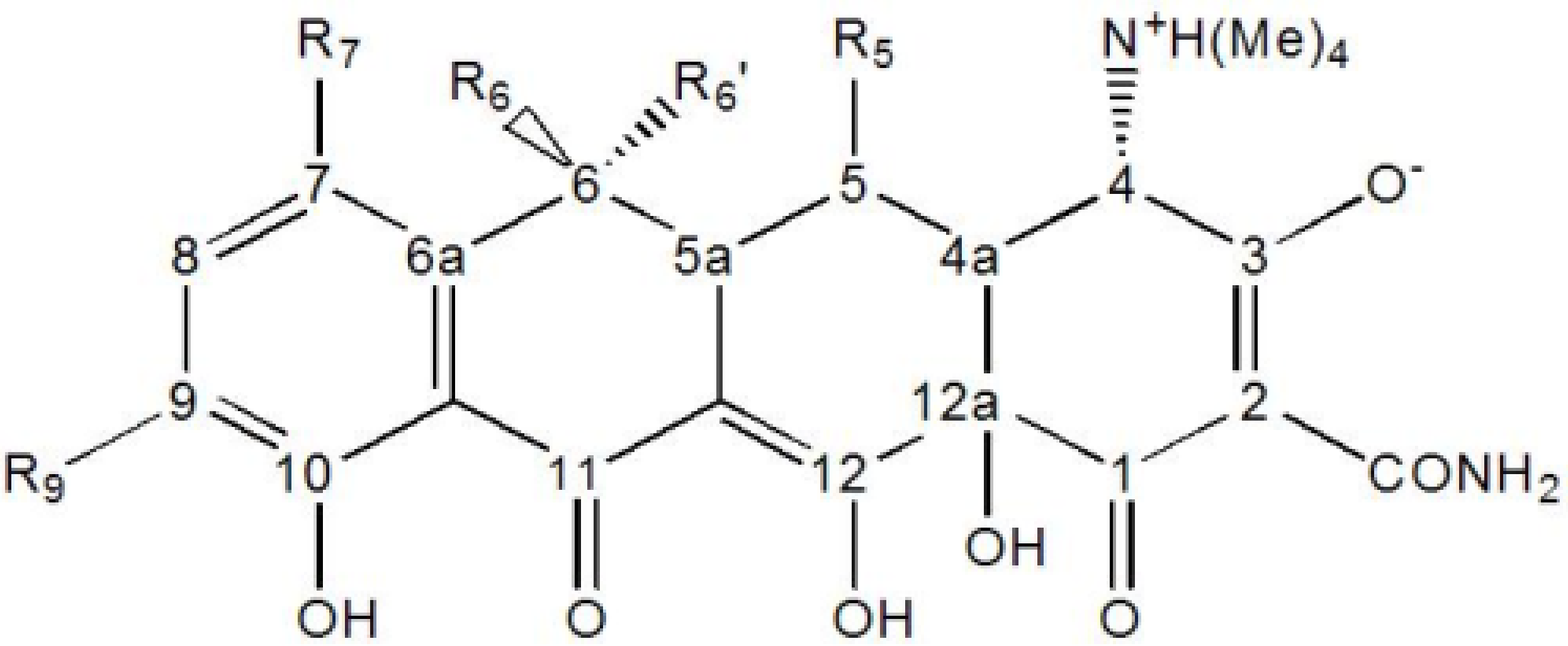}
\end{figure}

\begin{figure}
\caption{\label{fig5}Structures of new proposed tetracyclines - $T_{\beta}$}
\includegraphics[width=8.0cm]{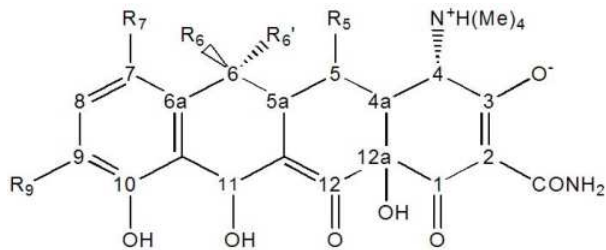}
\end{figure}

\begin{figure}
\caption{\label{fig6}Structures of new proposed tetracyclines - $T_{\gamma}$}
\includegraphics[width=8.0cm]{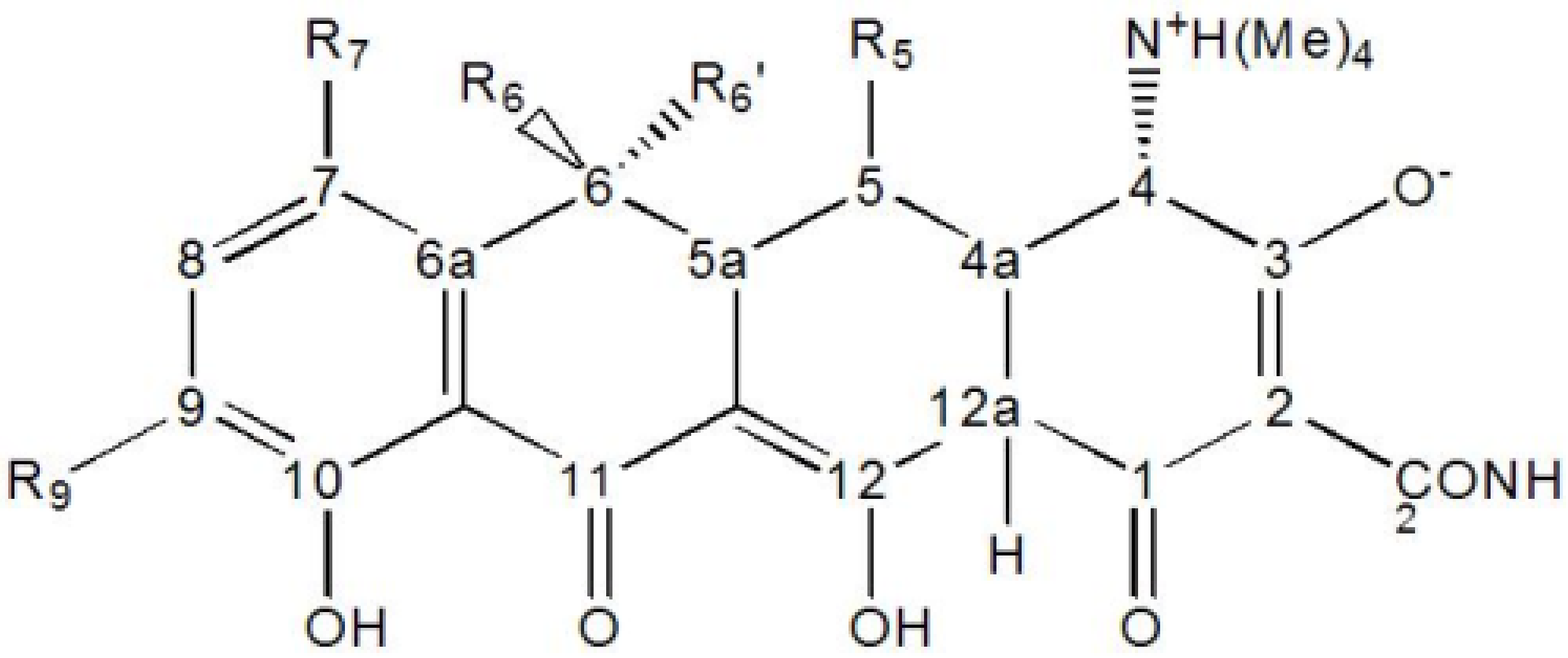}
\end{figure}

We considered the structures of molecules 01, 12, and 13 as our basic skeletons ($T_{\alpha}$, $T_{\beta}$, and $T_{\gamma}$) to generate new hypothetical TC (Fig. \ref{fig4}, \ref{fig5}, and \ref{fig6}). Combining these skeletons with common TC substituents (Table \ref{tab1}), plus some 7A column elements of the periodic table it was possible to generate 90 synthesis feasible new TC derivatives. Table \ref{tab3} presents the 30 combinations of the substituents considered for each skeleton. The new 90 compounds where labelled as 01 to 30, 31 to 60, and 61 to 90 for substituents of Table \ref{tab3} in $T_{\alpha}$, $T_{\beta}$ and $T_{\gamma}$, respectively.

\begin{table}
\caption{\label{tab3}New proposed TC compounds. Labelling accordingly to Fig. 7 and 8.}
\begin{tabular}{cccccc}
Structural set & R$_{5}$& R$_{6}$& R$_{6\mbox{'}}$& R$_{7}$& R$_{9}$ \\ \hline
01& Br& H& H& H& H \\
02& Br& H& H& H& Br \\
03& H& H& H& H& Br \\
04& H& H& Br& H& H \\
05& H& H& H& Cl& H \\
06& Cl& H& H& H& H \\
07& Cl& H& H& H& Cl \\
08& H& H& Cl& H& H \\
09& H& OH& H& Br& H \\
10& H& OH& Me& Br& H \\
11& H& OH& H& I& H \\
12& H& OH& Me& I& H \\
13& H& OH& H& F& H \\
14& H& OH& Me& F& H \\
15& I& H& H& H& H \\
16& I& H& H& H& I \\
17& H& H& H& H& I \\
18& H& H& H& I& H \\
19& H& H& I& H& H \\
20& H& H& H& F& H \\
21& H& OH& Me& NH$_{2}$& H \\
22& H& OH& Me& NO$_{2}$& H \\
23& H& OH& Me& H& NH$_{2}$ \\
24& H& OH& Me& H& NO$_{2}$ \\
25& OH& OH& OH& OH& OH \\
26& OH& H& H& H& N(Me)$_{2}$ \\
27& H& OH& Me& H& N(Me)$_{2}$ \\
28& H& H& H& OH& N(Me)$_{2}$ \\
29& OH& H& H& H& NH$_{2}$ \\
30& OH& H& H& H& NO$_{2}$ \\
\end{tabular}
\end{table}

These new compounds were then geometrically optimized following the same conformational searches  described in section \ref{metodologia} and for the global minima we similarly obtained the EIM parameters values ($\Delta H$ and $\eta H$) from group $CONH_{2}$ bound to carbon 2 (Table  \ref{tab5}) accordingly to equations \ref{deltah} and \ref{etah}. For these new compounds we considered only AM1 data, since the EIM results proved to be very similar using AM1 or PM3 and level of accuracy for AM1 was 100\%.

We use the EIM rules derived from AM1 calculations (equation \ref{eim-am1}) to classify the new TC as active or inactive (table \ref{tab5}). As can be observed from Table \ref{tab5}, 25 compounds are classified as actives while the remaining 66 are classified as inactives.

\begin{table*}
\caption{ \label{tab5}EIM descriptors calculated to the new proposed TC and the proposed biological qualitative activity (QA). I and A refer to the inactive and active, repectively.}
\begin{footnotesize}
\begin{tabular}{cccccccccccc}
Molecule& $\Delta$H& $\eta$H& QA& Molecule& $\Delta$H& $\eta$H& QA& Molecule & $\Delta$H& $\eta$H& QA  \\ \hline

01& 0.82&  1.39& I& 31& 0.80&  1.39& I& 61& 0.84&  1.42& I \\
02& 0.80&  1.39& I& 32& 0.90&  1.39& I& 62& 0.83&  1.42& I \\
03& 0.86&  1.39& I& 33& 0.93&  1.39& I& 63& 0.87&  1.42& I \\
04& 0.93&  1.39& I& 34& 0.87&  1.39& I& 64& 0.95&  1.42& I \\
05& 0.83&  1.39& I& 35& 0.84&  1.39& I& 65& 0.84&  1.42& I \\
06& 0.82&  1.39& I& 36& 0.81&  1.39& I& 66& 0.84&  1.42& I \\
07& 0.76&  1.39& I& 37& 0.89&  1.39& I& 67& 0.78&  1.42& I \\
08& 0.96&  1.39& I& 38& 0.89&  1.39& I& 68& 0.97&  1.42& I \\
09& 1.13&  1.23& A& 39& 1.04&  1.39& I& 69& 1.00&  1.42& I \\
10& 0.93&  1.39& I& 40& 0.91&  1.39& I& 70& 0.94&  1.42& I \\
11& 1.04&  1.38& A& 41& 1.03&  1.39& I& 71& 1.05&  1.41& I \\
12& 0.97&  1.39& I& 42& 0.94&  1.39& I& 72& 0.98&  1.42& I \\
13& 0.91&  1.39& I& 43& 0.93&  1.39& I& 73& 0.91&  1.42& I \\
14& 0.84&  1.39& I& 44& 0.84&  1.39& I& 74& 0.86&  1.42& I \\
15& 0.82&  1.39& I& 45& 0.79&  1.39& I& 75& 0.84&  1.42& I \\
16& 0.86&  1.39& I& 46& 0.90&  1.39& I& 76& 0.88&  1.42& I \\
17& 0.91&  1.39& I& 47& 0.93&  1.39& I& 77& 0.92&  1.42& I \\
18& 0.94&  1.39& I& 48& 0.92&  1.39& I& 78& 0.95&  1.42& I \\
19& 0.90&  1.39& I& 49& 0.85&  1.39& I& 79& 0.92&  1.42& I \\
20& 0.78&  1.39& I& 50& 0.81&  1.39& I& 80& 0.79&  1.42& I \\
21& 0.05&  1.39& A& 51& 0.16&  1.39& A& 81& 0.08&  1.42& A \\
22& 1.18& -0.26& A& 52& 1.02&  1.39& I& 82& 1.23&  0.28& A \\
23& 0.26&  1.39& A& 53& 0.43&  1.39& A& 83& 0.28&  1.43& A \\
24& 1.17& -0.50& A& 54& 1.19& -0.55& A& 84& 1.22& -0.44& A \\
25& 0.68&  1.39& I& 55& 0.83&  1.39& I& 85& 0.73&  1.43& I \\
26& 0.85&  1.39& I& 56& 0.86&  1.39& I& 86& 0.85&  1.42& I \\
27& 0.03&  1.39& A& 57& 0.19&  1.40& A& 87& 0.25&  1.43& A \\
28& 0.13& -1.39& A& 58& 0.03&  1.39& A& 88& 0.04&  1.42& A \\
29& 0.17&  1.39& A& 59& 0.34&  1.39& A& 89& 0.22&  1.43& A \\
30& 1.16& -0.38& A& 60& 1.17&  1.33& A& 90& 1.22& -0.19& A \\
\end{tabular}
\end{footnotesize}
\end{table*}

The EIM and PCA methodologies consider the investigation of the correlation of theoretical descriptors with biological experimental data through linear combinations of variables. To verify the behavior of EIM descriptors (data from Table \ref{tab2}) in non linear correlations with experimental data we performed property-activity relationship searches using Artificial Neural Network methodology (ANN) \cite{psdd1, psdd2}.
 
For ANN investigations we considered the 14 molecules studied in section III-A as the training group and the EIM parameters as network variables. In this trainning pattern we required the interactions precision of $10^{-5}$ and a $3\times10^{5}$ of interactions. In Table \ref{tab6} we present the parameters used in ANN calculations, where $\alpha$ is a nonlinearity parameter of sigmoid functions of neurons in the second and third neuron's layer, $\theta$ is a threshold neuron parameter in a second and third layers, and $\epsilon$ is a constant weight mixing the weight matrix between recursive cicles.  

\begin{table}
\caption{\label{tab6}Parameters of ANN for training pattern.}
\begin{tabular}{ccccc}
Layer & Neurons & $\alpha$ & $\theta$ & $\epsilon$ \\ \hline
1     & 3       &        &        &     \\
2     & 20      & 3.0    & 0.0    & 0.2 \\
3     & 2       & 3.0    & 0.0    & 0.2 \\
\end{tabular}
\end{table}

In Table \ref{tab7} we present the results for the training with the 14 molecules. The input data was set as class 1 to active molecules (1.000 0.000 - pattern) and class 2 to inactive molecules (0.000 1.000 - pattern). The columns four and five present the output values of training compared to the 0.000 and 1.000 values of input. We can observe that output values are very close to input, indicating a good training of the network. This good learning was achieved with precision parameters in cycle 242,477.

\begin{table}
\caption{\label{tab7}Input and output pattern of ANN training using EIM index of AM1 calculations of table \ref{tab2}.}
\begin{tabular}{cccc}
No. & Trainning Pattern & Output Trainning & Class \\ \hline
01& 1.000  0.000& 0.996  0.004& 1 \\
02& 1.000  0.000& 1.000  0.000& 1 \\
03& 1.000  0.000& 1.000  0.000& 1 \\
04& 1.000  0.000& 1.000  0.000& 1 \\
05& 1.000  0.000& 1.000  0.000& 1 \\
06& 1.000  0.000& 1.000  0.000& 1 \\
07& 1.000  0.000& 0.997  0.003& 1 \\
08& 1.000  0.000& 1.000  0.000& 1 \\
10& 1.000  0.000& 1.000  0.001& 1 \\
11& 1.000  0.000& 1.000  0.000& 1 \\
12& 1.000  0.000& 1.000  0.000& 1 \\
09& 0.000  1.000& 0.000  1.000& 2 \\
13& 0.000  1.000& 0.005  0.995& 2 \\
14& 0.000  1.000& 0.003  0.997& 2 \\
\end{tabular}
\end{table}

In Table \ref{tab8} we present the results of the ANN classification, classes 1 and 2 correspond to active and inactive compounds, respectively. For comparision the EIM classification is also displayed. The results of EIM and ANN are in very good agreement. Only 6 compounds (11, 12, 39, 41, 52, and 71) out of $\sim$90\% (agreement) present conflictive classification. Interestingly all these 6 molecules are classified as active by ANN and inactive ones by EIM, which seem to have more restrictive conditions for active classification.

\begin{table*}
\caption{\label{tab8} Results of the ANN prediction to the new TC compounds using only EIM parameters. Classes 1 and 2 indicate active and inactives compounds respectively}
\begin{footnotesize}
\begin{tabular}{ccccccccccccccc}

& \multicolumn{3}{c}{ANN} & EIM & & \multicolumn{3}{c}{ANN} & EIM & & \multicolumn{3}{c} {ANN} & EIM \\ \hline
No. & \multicolumn{2}{c}{Predict} & Class & Class & No. & \multicolumn{2}{c}{Predict} & Class & Class & No. & \multicolumn{2}{c}{Predict} & Class & Class \\
1& 0.002& 0.998& 2& 2& 31& 0.001& 0.999& 2& 2& 61& 0.000& 1.000& 2& 2 \\
2& 0.001& 0.998& 2& 2& 32& 0.017& 0.983& 2& 2& 62& 0.000& 1.000& 2& 2 \\
3& 0.001& 0.999& 2& 2& 33& 0.015& 0.986& 2& 2& 63& 0.000& 1.000& 2& 2 \\
4& 0.044& 0.959& 2& 2& 34& 0.001& 0.999& 2& 2& 64& 0.000& 1.000& 2& 2 \\
5& 0.001& 0.999& 2& 2& 35& 0.000& 0.999& 2& 2& 65& 0.000& 1.000& 2& 2 \\
6& 0.002& 0.998& 2& 2& 36& 0.001& 0.999& 2& 2& 66& 0.000& 1.000& 2& 2 \\
7& 0.001& 0.999& 2& 2& 37& 0.007& 0.993& 2& 2& 67& 0.000& 1.000& 2& 2 \\
8& 0.383& 0.633& 2& 2& 38& 0.002& 0.998& 2& 2& 68& 0.000& 1.000& 2& 2 \\
9& 1.000& 0.000& 1& 1& 39& 1.000& 0.000& 1& 2& 69& 0.087& 0.911& 2& 2 \\
10& 0.066& 0.938& 2& 2& 40& 0.005& 0.995& 2& 2& 70& 0.000& 1.000& 2& 2 \\
11& 1.000& 0.000& 1& 1& 41& 1.000& 0.000& 1& 2& 71& 1.000& 0.000& 1& 2 \\
12& 0.924& 0.080& 1& 2& 42& 0.032& 0.970& 2& 2& 72& 0.002& 0.998& 2& 2 \\
13& 0.016& 0.984& 2& 2& 43& 0.021& 0.980& 2& 2& 73& 0.000& 1.000& 2& 2 \\
14& 0.001& 0.999& 2& 2& 44& 0.001& 0.999& 2& 2& 74& 0.000& 1.000& 2& 2 \\
15& 0.002& 0.998& 2& 2& 45& 0.001& 0.999& 2& 2& 75& 0.000& 1.000& 2& 2 \\
16& 0.005& 0.995& 2& 2& 46& 0.015& 0.985& 2& 2& 76& 0.000& 1.000& 2& 2 \\
17& 0.007& 0.993& 2& 2& 47& 0.015& 0.986& 2& 2& 77& 0.000& 1.000& 2& 2 \\
18& 0.138& 0.870& 2& 2& 48& 0.004& 0.996& 2& 2& 78& 0.000& 1.000& 2& 2 \\
19& 0.006& 0.994& 2& 2& 49& 0.001& 0.999& 2& 2& 79& 0.000& 1.000& 2& 2 \\
20& 0.000& 1.000& 2& 2& 50& 0.000& 1.000& 2& 2& 80& 0.000& 1.000& 2& 2 \\
21& 1.000& 0.000& 1& 1& 51& 1.000& 0.000& 1& 1& 81& 1.000& 0.000& 1& 1 \\
22& 1.000& 0.000& 1& 1& 52& 1.000& 0.000& 1& 2& 82& 1.000& 0.000& 1& 1 \\
23& 1.000& 0.000& 1& 1& 53& 1.000& 0.000& 1& 1& 83& 1.000& 0.000& 1& 1 \\
24& 1.000& 0.000& 1& 1& 54& 1.000& 0.000& 1& 1& 84& 1.000& 0.000& 1& 1 \\
25& 0.004& 0.995& 2& 2& 55& 0.001& 0.999& 2& 2& 85& 0.000& 1.000& 2& 2 \\
26& 0.002& 0.998& 2& 2& 56& 0.001& 0.999& 2& 2& 86& 0.000& 1.000& 2& 2 \\
27& 1.000& 0.000& 1& 1& 57& 1.000& 0.000& 1& 1& 87& 1.000& 0.000& 1& 1 \\
28& 1.000& 0.000& 1& 1& 58& 1.000& 0.000& 1& 1& 88& 1.000& 0.000& 1& 1 \\
29& 1.000& 0.000& 1& 1& 59& 1.000& 0.000& 1& 1& 89& 1.000& 0.000& 1& 1 \\
30& 1.000& 0.000& 1& 1& 60& 1.000& 0.000& 1& 1& 90& 1.000& 0.000& 1& 1 \\
\end{tabular}
\end{footnotesize}
\end{table*}

Finally, it is interesting to note the role played by substituents in the biological behavior of the new set of tetraciclines. According to our qualitative analysis the main substitution requirement for biological activity is the presence of NH$_{2}$ or NO$_{2}$ or dialkylamino groups at C7 and/or C9. Even for non-active derivative 13 (Table I) the presence of these substituents classifies as active new compounds 81-84 and 87-90 (Table VIII). Despite the whole molecular skeleton composed of 4 fused rings and the integrity of the ring A, the rest of the molecule including position C5 and C6 just plays minor role on the biological activity of tetracyclines.

\section{Summary and Conclusions}\label{conclu}

In this work we applied the EIM methodology in the classification of activity of 14 derivatives of the Tetracicline and in the predictive classification of the activity of 90 hypothetical new derivatives. Our results showed that the EIM results are not method dependent indicating that the classification of activity of Tetraciclines is reproducible for different semiempirical methods. In comparison with conventional physicochemical descriptors the importance of EIM descriptors is reinforced by PCA and ANN methodologies. The rules constructed with EIM, PCA and ANN methods reproduced the experimental data with 100\%, 96\% and 100\% of accuracy, respectivelly. For the new 90 derivatives proposed EIM and ANN are in good agreement with a percentage of 93\% consistent classification. These results add to increasing data of EIM studies indicating the pure quantum electronic indices can be reliably used to classify the biological activity of large number of different classes of materials.

We hope the present study estimulate further experimental studies for tetracyclines in order to try to produce new antibiotic compounds.

\section{Acknowledgements}

This work has been supported by the Brazilian agency FAPESP and also thanks to CAPES, CNPq, FAPEMIG, and IMMP.


\end{document}